\documentstyle[prl,aps,epsf]{revtex}
\begin{document}
\draft

\twocolumn[\hsize\textwidth\columnwidth\hsize\csname %
@twocolumnfalse\endcsname

\title
{d-Wave Pairing Correlation in the Two-Dimensional $t-J$ Model}

\author{ C.T. Shih$^1$, Y.C. Chen$^2$, H. Q. Lin$^3$, and T.K. Lee$^4$}
\address{
$^{1}$National Center for High-Performance Computing, Hsinchu, Taiwan\\
$^{2}$Dept. of Physics, Tunghai Univ., Taichung, Taiwan\\
$^{3}$Dept. of Physics, Chinese Univ. of Hong Kong, Hong Kong\\
$^{4}$
Inst. of Physics, Academia Sinica, Nankang, Taipei, Taiwan
}

\date{\today}
\maketitle
\begin{abstract}
The pair-pair correlation function of the two-dimensional $t-J$ model
is studied by using the power-Lanczos method and an assumption of
monotonic behavior.
In comparison with the results
of the ideal Fermi gas, we conclude that the 2D $t-J$ model does {\it not}
have long range d-wave superconducting  correlation in the 
 interesting parameter range of $J/t \leq 0.5$.
Implications of this result will also be discussed.

\end{abstract}
]

It is believed that
the two-dimensional (2D) $t-J$ model is a reasonable starting point to 
understand the physical properties
of the high temperature superconducting (HTSC) cuprates
\cite{anderson87,zhang88}. One of the 
critical issues is whether  
the model has enough ingredients to quantitatively explain the
high $T_c$.
Here we shall report a numerical study to address this issue.

 The 2D $t-J$ model on a square lattice is:
\begin{equation}
H=-t\sum_{<i,j>\sigma} (\tilde{c}^\dagger_{i\sigma}\tilde{c}_{j\sigma}+h.c.)
  + J\sum_{<i,j>}({\bf S}_i\cdot{\bf S}_j-{1\over4} n_in_j),
\end{equation}
where $\tilde{c}^\dagger_{i\sigma} = c^\dagger_{i\sigma}(1-n_{i-\sigma})$, and
$\langle i,j\rangle$ denotes the nearest neighbors $i$ and $j$.
Using the
experimental results of HTSC, we expect the physical interesting 
value of $J/t$ to be about $0.4$ and that
superconductivity should occur for electron density $n_e$ greater than
$0.7$.

The first indication about the superconductivity of the $t-J$ model
is to determine if the two holes would form a bound state by
 the exact diagonalization study on small lattices
\cite{poilblanc94}. 
However, the attractive potential among doped holes is not
a sufficient condition for superconductivity. 
 The long-range pair-pair correlation  should be a better indicator.
The pair-pair correlation function is defined as:
\begin{equation}
P_{s\ or\ d}(R)=\frac{1}{N_s} \sum_{i} \langle \Delta^\dagger_i
\Delta_{i+R}\rangle
\label{e:pairing}
\end{equation}
where
$\Delta_i = c_{i\uparrow}(c_{i+\hat{x}\downarrow}
   + c_{i-\hat{x}\downarrow} \pm c_{i+\hat{y}\downarrow}
   \pm c_{i-\hat{y}\downarrow})$. $+$ and $-$ represent  $s$ and
$d_{x^2-y^2}$ pairing states respectively.
$N_s$ is the number of sites.

White {\it et al.} \cite{white89} studied the one-band
Hubbard model and suggested that 
at low temperatures the pair-field
susceptibility $\chi_d = \sum_R P_d(R)$ is enhanced 
in the $d_{x^2-y^2}$ 
channel 
 and  small for others.
Other convincing results are from variational Monte Carlo studies\cite{gros88}.
Although it did not provide quantitative values for $T_c$, the magnitude
of pair-pair
correlation varies with the hole density in a similar way as $T_c$ of HTSC.
 The prediction of the dominance of d-wave pairing, instead of s-wave,
 in the $t-J$ model is
also quite encouraging.

However, these results are not quite consistent with a recent report by 
 Zhang {\it et al.}. 
They studied the 2D one-band\cite{zhang97} 
and three-band Hubbard model\cite{zhang98}
by using the constrained path Monte Carlo (CPMC) method.
They concluded that for $U/t > 4$ the long-range pair-pair
correlation vanishes. 
 It becomes quite important
to  have a careful numerical study of 
the pairing correlation in
the 2D $t-J$ model for larger lattices. In this paper
we show that it is likely that there is no
long-range $d_{x^2-y^2}$ pair-pair correlation at all
for the {\it two-dimensional} $t-J$
model
 in the physical parameter range
($J/t \leq 0.5$).

In the variational level, the optimal state of the 2D $t-J$ model
for a range of parameters is 
the $d_{x^2-y^2}$ RVB trial wave function:
\begin{equation}
\mid RVB \rangle = P_d \prod_k (\tilde{u}_k +
\tilde{v}_k c^\dagger_{k,\uparrow}
c^\dagger_{-k,\downarrow})\mid 0 \rangle
\label{e:TWRVB}
\end{equation}
with
$\tilde{v}_k/\tilde{u}_k =
\Delta_k/(\epsilon_k + \sqrt{\epsilon_k^2
+\Delta_k^2})$,
 $\Delta_k = \Delta(cosk_x - cosk_y)$ and
$\epsilon_k = 2(cosk_x + cosk_y) - \mu$.
 $\Delta$ is the d-wave superconducting order parameter
and $\mu$ is the chemical potential. The operator $P_d$ enforces the
constraint of no double occupancy. We take $t=1$ in this paper.
This wave function with the form of a projected BCS wave function is
known to be superconducting \cite{gros88}.

It is well known that the variational calculation usually
overestimates the effect of superconductivity of the true ground state.
For the $t-J$ model  
the energy is dominated by the nearest neighbor interaction.
Hence it naturally  leads to large $\Delta$ in the variational calculation.   
Heeb and Rice \cite{heeb94} suggested that to examine the true
pairing correlation, it might not be a good idea to use the lowest
variational energy as a criterion in selecting the parameters of
the trial function.
They modified the function Eq.({\ref{e:TWRVB}) with parameters 
that they believe 
 can separate the short-  and long-range
parts of the correlations. 
They found the critical $J_c\simeq0.44$ for the
onset of superconducting long-range order 
 for $n_e=42/50$.
 Here we modify their idea 
with a more systematic approach and provide with more rigorous analysis.

One of the ways to eliminate  the bias induced by the trial wave
function is to project the trial function onto the true ground state of
the system. 
Recently we have shown that
the ground state energy and many other properties might be obtained accurately
by using a particular ground state projection method, the
 power-Lanczos method
\cite{yctk95,ctyctk97}, which is a hybrid of the power method and the
variational Lanczos method.
In the power method it can be easily shown that
if a trial wave function $| \Psi \rangle$ is not orthogonal to the 
ground state,  $(W-H)^m \mid \Psi \rangle$ is 
proportional to the ground state wave function as the power $m$ approaches
infinity. $W$ is an appropriately chosen constant to make the ground 
state energy the largest eigenvalue of the $W-H$ matrix.
 However in practice, due to the Fermion sign problem the power 
cannot be too large. A better way is to 
improve the trial function $\mid \Psi \rangle$
by using the first order Lanczos method, i.e., to use 
$\mid PL1 \rangle = (1+C_1 H)\mid \Psi \rangle$ 
and then apply the power of $W-H$.
$C_1$ is a new variational parameter. The results described below are either
calculated by  $\mid \Psi \rangle$, denoted by PL0-V, or by 
 $(W-H)^m \mid PL1 \rangle$ as PL1-Pm. $m=0$ is the 
variational result, PL1-V, of the $\mid PL1 \rangle$ state.

\begin{figure}[m]
\epsfysize=6cm\epsfbox{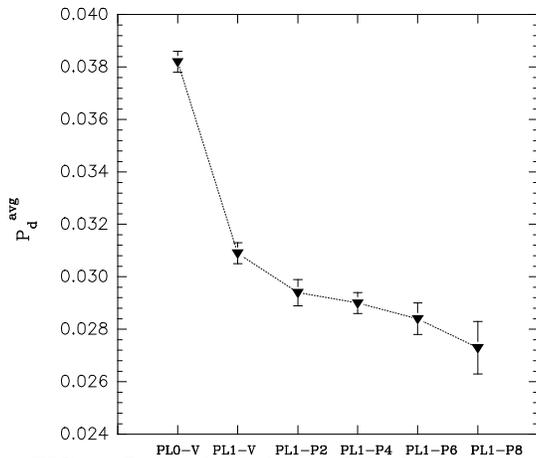}
\caption{Power dependence of the long-range pairing average
$P_d^{avg}$ for $n_e=42/50$,
$\Delta=0.64$, and $J=0.7$.}
\label{f:pwd32}
\end{figure}

Clearly, if the trial function is a very good representation of the 
ground state, the various correlation functions 
calculated in PL0-V should be almost the same
as in PL1-Pm. On the other hand, if the trial function is biased in the
wrong way, results of PL1-Pm will be very different from PL0-V and it 
will correct the bias\cite{yctk95,ctyctk97}. This is demonstrated in
 Fig.\ref{f:pwd32}.

The averaged value, $P_d^{avg}$, of the
long-range part ($R > 3$) of $P_d(R)$ is plotted as a function of powers
in Fig.\ref{f:pwd32} for $\Delta=0.64$ which gives the lowest
variational energy.
The electronic density is $n_e=42/50$ and 
$J=0.7$. The value of $P_d^{avg}$ is suppressed substantially from
the VMC or PL0-V result when the power is increased. 
Clearly the optimized trial function has grossly 
overestimated the strength
of the long range pair-pair correlation. As noted above, this is 
due to the choice of a large $\Delta$ to optimize the short-range pairing.

Although the
result in Fig.\ref{f:pwd32} 
is not yet converged, we could see that the 
 long-range pair correlation changes 
{\it monotonically} as the wave function approaches the ground state. 
A more clear demonstration of this monotonic behavior is shown in Fig.\ref{f:HRpwpair}.

\begin{figure}[m]
\epsfysize=6cm\epsfbox{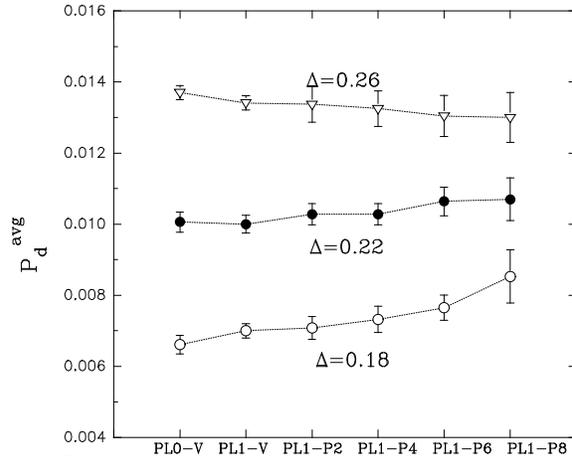}
\caption{Power dependence of the long-range pairing average
$P_d^{avg}$ for several trial wave functions with different
$\Delta$ for $n_e=42/50$ and $J=0.7$.}
\label{f:HRpwpair}
\end{figure}

 $P_d^{avg}$ is plotted as a function of powers 
in Fig.\ref{f:HRpwpair} for three different values of $\Delta$: 
open circles are for $\Delta=0.18$, full circles for $0.22$ and
triangles for $0.26$. 
$P_d^{avg}$ remains almost unchanged for $\Delta=0.22$. For 
$\Delta$ larger than $0.22$, the pair correlation always decreases with
power. And the opposite is true for
$\Delta$ smaller than $0.22$. Since $P_d^{avg}$ for $\Delta=0.22$
 hardly changes
as the state gets closer and closer to the ground state, it is natural
to assume that this is the ground state result. Moreover, if this is
a good criterion to determine the ground state value, then we really 
only need to calculate PL0-V, PL1-V and PL1-P2. There is no need to
go to larger powers and the minus sign problem is avoided. 
The same result would be obtained if we examine $P_d(R)$ for the
largest $R$ instead of using $P_d^{avg}$.

In addition to $\Delta$, $\mu$ is also a variational parameter in the
RVB wave function. 
Just like $\Delta$ which is not the real
superconducting order parameter, 
$\mu$ is not the real chemical potential as in the simple
mean field equations. 
In the discussion in the previous
paragraph, $\mu$ is chosen to be consistent with the Fermi
surface of the ideal fermi gas. For example, $\mu=-0.4$ for $n_e=42/50$.
If $P_d^{avg}$ is the true ground state value, then just like 
$\Delta$ no matter what
initial values of $\mu$ we use for the trial wave function, the
converged result would remain the same. In other words, 
different sets of ($\Delta$, $\mu$) will converge to the same 
final $P_d^{avg}$. This important consistency check 
has been verified. 
For example, for $n_e=42/50$, we obtain  $P_d^{avg}=0.0245(4)$ 
for $\Delta_{J=1}=0.4$ and $\mu=-0.4$;  $P_d^{avg}=0.0238(4)$ 
for $\Delta_{J=1}=0.34$ and $\mu=-0.6$;  and  $P_d^{avg}=0.0238(3)$ 
for $\Delta_{J=1}=0.24$ and $\mu=-0.8$. 

So far by using the ground state projection method,
we have shown that  
the long-range pair correlation
$P_d^{avg}$ 
  approaches  the ground-state
value  monotonically. This basic assumption is used   
   to choose parameters
to best represent the ground-state value of $P_d^{avg}$
 instead of determining variational 
parameters according to the lowest energy criterion. 
Unfortunately there are no exact calculations 
for the 2D $t-J$ model with large lattices to
test our assumption. However, the method could be tested in the study of the
pairing correlation for the 2D attractive and repulsive Hubbard models.

\begin{figure}[m]
\epsfysize=5cm\epsfbox{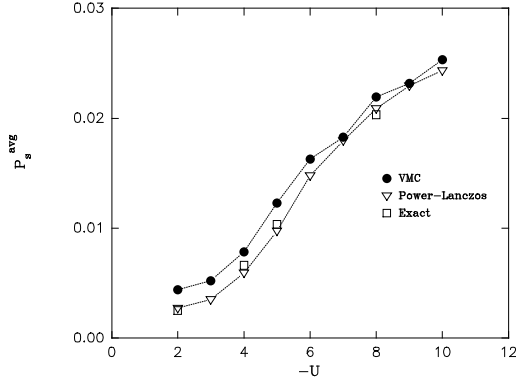}
\caption{Pairing correlation as a function of  $U$ for $n_e=4/64$.
Full circles are results
evaluated from the energy-optimizing
trial wave functions, open squares for exact results and open triangles
are results based on our method.
}
\label{f:ahu0464}
\end{figure}
It is known that 
the  2D attractive
Hubbard model
has long-range s-wave pairing correlation.
The on-site pairing correlation, $\Delta_i = c_{i\uparrow}c_{i\downarrow}$,
for several $U$ is shown in Fig.\ref{f:ahu0464}. We consider the electron
density at 4/64 which is solved exactly. The figure shows that
the energy-optimizing trial wave functions (full circles) always
overestimate the pairing correlation in comparison with the
exact results (open squares). And the results 
obtained by our power-Lanczos method 
 (open triangles) are in much better agreement.

We have examined the long-range d-wave
 pairing correlation for the 2D repulsive Hubbard model. Our result
agrees with 
Zhang et al\cite{zhang97} that the long-range pairing correlation
is vanishingly small.   

The success for the attractive and repulsive Hubbard models further support
our method. The method allows us to 
calculate the ground state  $P_d^{avg}$ for large lattices. 
In Fig.\ref{f:prj5}, 
$P_d^{avg}$ is plotted a function of electron density for 
82 and 122 sites with $J=0.5$. 
The PL0-V and PL1-V results for the
trial wave functions with $\Delta_{opt}$ optimizing the variational
energies are also shown in the insets of Fig.\ref{f:prj5} for 82 sites.
It is clear that $P_{d,opt}^{avg}$
is substantially reduced by PL1.
And the variational values are an order of magnitude greater
than the ground state values. 

Further, the $P_d^{avg}$ for $J=0.5$ can be compared with that of the
ideal fermi gas (IFG). 
The error bars denote the range of values for different degenerate
states of the IFG.
The nonzero $P_d^{avg}$ 
is obviously a finite size effect.
Since  $P_d^{avg}$ for
$J=0.5$ are smaller than that of the IFG we can 
easily conclude
that there is no long range d-wave pairing correlation for $J=0.5$.

\begin{figure}[m]
\epsfysize=6cm
\epsfbox{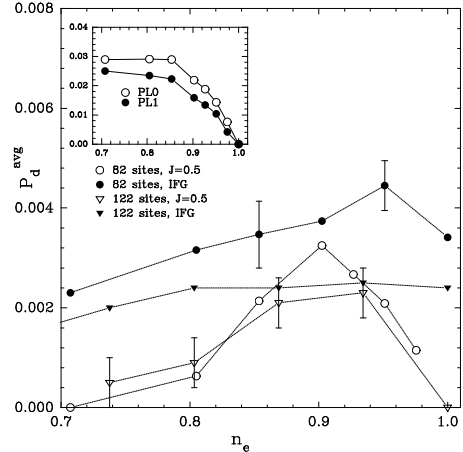}
\caption{$P^{avg}_d$ for $J/t=0.5$ and ideal fermi gas for
82 and 122 sites. The
PL0 and PL1 results of energy-optimizing $\Delta_{opt}$ for 82 sites
are shown in the
insets.
} \label{f:prj5}
\end{figure}

\begin{figure}[m]
\epsfysize=6cm
\epsfbox{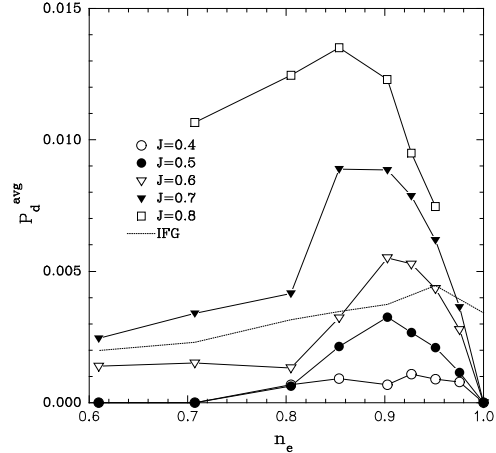}
\caption{$P^{avg}_d$ for 82 sites for different density and $J$.
} \label{f:pair82}
\end{figure}

In Fig.\ref{f:pair82} we show $P_d^{avg}$ for different densities and
$J$ for 82 sites. The dotted line is the result of the IFG. 
It can be seen that $P_d^{avg}$ is larger than the IFG values only for
$J \geq 0.6$, which is considerably larger than the physical value
of $J=0.3$ or $0.4$. The large values observed for $J\geq0.6$ could be
due to the phase separation\cite{ctyctk97}. 
 For $J\leq0.5$ 
not only 
$P_d^{avg}$ seems to be too small to give large $T_c$ for the HTSC materials,
the maximum $P_d^{avg}$ are at hole density $x_h=1-x_e\sim 10\%$. This
 disagrees with experiments which have 
optimal doping at $x_h\sim 15\%\sim 20\%$. For $J\geq0.6$,
the values of $x_h$ are very close to the  critical hole densities of phase
separation
\cite{ctyctk97}. 
It is possible that for large $J$ the $P_d^{avg}$ measured actually indicates
 electron clustering near phase separation rather than
superconductivity.

Our result that there is no long range
d-wave pairing correlation  for $J\leq0.5$
 is actually consistent with the exact 
numerical results for the two-hole binding energy. It is known that 
binding two holes is a necessary condition for the occurrence of
superconductivity.
In Fig.\ref{f:bind}
 the binding energy of two holes for various J
is plotted 
as a function of the inverse of $N_s$ which is the cluster size.
The results of 32 sites are obtained by Leung\cite{leung98}
and 26 sites by Poilblanc\cite{poil93}. It shows that 
two holes do not bind together unless $J>0.8$.
Due to the different cluster shapes, the data are not quite on a straight line.
However, even taking into account the deviation, the result
is still consistent with the absence of hole
binding for $J\leq0.5$.

\begin{figure}[m]
\epsfysize=6cm
\epsfbox{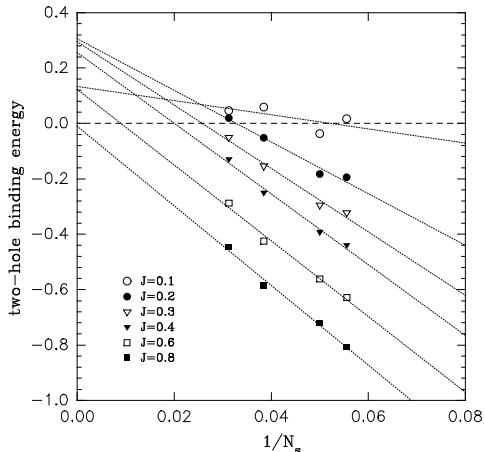}
\caption{Two-hole binding energies as a function of the inverse of cluster
size for different $J$.}
\label{f:bind}
\end{figure}

In summary, based on a simple observation 
that the long range d-wave pairing correlation changes monotonically
when the state approaches the ground state, we assume that 
the ground state value can be determined by choosing parameters that do not
optimize the variational energy but keeps correlation value unchanged
when we project the wave function onto the ground state. As a consistency
check we show that different
sets of values of parameters,  $\Delta$ and $\mu$, produce same 
correlation value. 
This method has also been successfully tested for the 2D attractive and
repulsive Hubbard models. The result that the long range d-wave 
pairing correlation is nonzero only for $J\geq0.6$ is consistent 
with the finite size analysis of exact 
results for small clusters. It is also consistent with
the result obtained in a completely different analysis that the phase
separation boundary occurs only for  $J\geq0.6$. 

Since there are a number of evidences that the 2D t-J model
is a fairly good model for HTSC, the negative result with respect to
the d-wave pairing correlation reported above is quite disturbing.
There are several possibilities to explain this discrepancy.  

The first thing one can point to is the possible contribution of
next-nearest neighbor hopping and next-next-nearest neighbor hopping,
 $t'$ and $t''$ respectively,
to the density of states. By choosing appropriate 
 $t'$ and $t''$,  the density of states may have van Hove singularities
at hole density around $15\%$. This might greatly enhance
superconductivity.
Contrary to this belief we found that
the superconductivity is not enhanced by adding $t'$ term
 even though we have tuned the parameters to
have the Fermi surface passing through the van Hove singularity
at  $15\%$ hole density. Details will be reported elsewhere\cite{ctyctk97b}. 

Another possibility is that due to other interactions such as electron-phonon
the effective $J/t$ might be larger than
$0.6$. However in this case, it is more likely we will have phase separation.
The doping dependence of $T_c$ is also inconsistent with experiments.
The interlayer tunneling model proposed by Anderson et al
\cite{bilayer}  certainly expects the absence of pairing in our pure 2D model.
More exotic possibility might be that the true ground state symmetry
is not d-wave but a time-reversal-broken order parameter\cite{ong,laughlin}.

We would like to thank P.W. Leung for giving us his data prior to
publication.
This work is partially supported by the National Science Council of
Republic of China, Grant Nos. NSC 86-2811-M-007-001R \&
87-2112-M-001-016 \& 86-2112-M-029-001; and the 
Research Grants Council (RGC) of the Hong
Kong
Government under account Nos. 2160089. \& 44M6003(CUHK).
Part of computations were performed at the National Center
for High-Performance Computing in Taiwan. We are grateful for
their support.
TK thanks Physics Department at CUHK for their hospitality.


\end{document}